\documentclass[11pt,fleqn]{article}
\usepackage{graphicx}
\usepackage{latexsym}
\usepackage{amssymb}
\usepackage{amsmath,amsthm}
\title{An Equation of Motion with Quantum Effect in
Spacetime \footnote{The author is partially supported by a Taiwan
NSC grant.}}
\author{\small{Jyh-Yang Wu\footnote{email:jywu@math.ccu.edu.tw}} \\
 \scriptsize {Department of Mathematics , National Chung Cheng University,}\\
  \scriptsize{Ming-Hsiung, Chia-Yi 621,
 Taiwan}}
 \date{}

\begin{document}
 \maketitle

\abstract In this paper, we shall present a new equation of motion
with Quantum effect in spacetime. To do so, we propose a
classical-quantum duality. We also generalize the Schordinger
equation to the spacetime and obtain a relativistic wave equation.
This will lead a generalization of Einstein's formula $E=m_0c^2$ in
the spacetime. In general, we have $E=m_0c^2 +
\frac{\hbar^2}{12m_0}R$ in a spacetime.

\section{Introduction}

The effort to combine the theories of General Relativity and Quantum
Mechanics can be tracked to the very beginning of these two
theories. Since the middle of this century, the study of Quantum
gravity has attracted a lot of attention. The usual way to combine
these two theories is to start from Quantum Mechanics (QM), via
Special Relativity (SR), and then reach General
Relativity (QR). This gives\\
\par Physical Approach (I) :
\par~~~~Quantum Mechanics $\mapsto$ Special Relativity $\mapsto$
General Relativity. \\

Following this path, one can obtain the celebrated Dirac equation
and the Klein-Gordon equation. However, the advance to the level of
General Relativity along this line has not yet succeeded. In other
words, one now has more knowledge about the quantum effect in
Special Relativity, but is not able to combine the quantum effect
with the general relativity effect, i.e., the quantum effect in
spacetime.

To unify the theories of Quantum Mechanics and General Relativity
one need to take care of three major effects : quantum effect,
(special) relativity effect and gravitation effect. The gravitation
effect is also called the general relativity effect which is an
effect of the curvedness of the spacetime. This physical approach
from Quantum Mechanics, via Special Relativity, to General
Relativity tries to catch the quantum-relativity effect first and
then try to combine it with the gravitation effect. In this note we
try to propose another way from Quantum Mechanics to General
Relativity, via Riemannian Geometry (RG). Namely, we
introduce\\
\par Geometric Approach (II) :
\par~~~~Quantum Mechanics $\mapsto$ Riemannian Geometry $\mapsto$
General Relativity.\\

This approach tries to catch the quantum-gravitation effect first
and then combine it with the relativity effect. Thus, the first
approach is more physical, while the second more geometric. A common
fundamental problem in both physics and geometry is to deal with the
motion of objects in a suitable underlying space. This is why there
are so many interactions between these two fields.

In order to illustrate the importance of the Geometric Approach
(II), we shall point out, in this note, a new equation of motion
with quantum effect in spacetime. To demonstrate why the geometry
plays an important role in physical world, we would like to point
out that both theories of Quantum Mechanics and General Relativity
are, in some respects, geometric. On one hand, it is well-known that
in Einstein's theory of General Relativity the gravitation is viewed
as the curved effect of spacetime. In other words, General
Relativity uses the curved spacetime to unify space, time and
gravity. Thus, General Relativity is geometric. On the other hand,
Feynman's path integral formulation of the theory of Quantum
Mechanics provides us a sum-over-all-path thinking. This amounts to
the consideration of all possible trajectories of a particle. Hence
this viewpoint is also geometric. These Observations leads us to
realize the important role of geometry that many possibly play in
these two theories and also give us a sufficient hint to put the
theory of Riemannian Geometry as the intermediate step from Quantum
Mechanics to Geometry Relativity even through at the first sight,
the theory of Riemannian Geometry may seem a little bit exotic from
the viewpoint of physics.
\par In section two we shall introduce a generalized Newton's
equation of motion. For a massive particle with rest mass $m_{o}$,
the timelike trajectory is governed by the generalized Newton's
equation of motion:
 \begin{equation} F =
m_{o}\nabla_{\gamma^{'}(\tau)}\gamma^{'}(\tau)-
\displaystyle\frac{\hbar^{2}}{12m_{o}}\nabla R
\tag{1.1}\end{equation} where $\nabla R$ is the gradient of the
scalar curvature $R$ as given by (2.2), $h$ the Planck constant and
$F$ the external force. For a massless particle, like photons, the
trajectory is given by the equation :

\begin{equation}
\nabla_{\gamma^{'}(\tau)}\gamma^{'}(\tau)=\displaystyle\frac{\lambda^{2}c^{2}}{48\pi^{2}}\nabla
R\tag{1.2}\end{equation}
without external force where $\lambda$
stands for the wavelength of the massless particle.
\par The Equation (1.2) does not contain the quantum parameter
$\hbar$. However, it comes from the consideration of quantum effect
on curved space-time. Moreover, this equation will give us a new
phenomenon of ``gravitational rainbow'' and also provide a new
explantation of the evaporation of a black hole rather than that
given by Hawking. See [Wu3-4] for detailed discussion of these new
effects.

In section three we give a detailed explanation how this equation
arises naturally. In section four we shall also develop a new
relativistic wave equation:

\begin{equation}
i\hbar\displaystyle\frac{\partial\Phi(x,\tau)}{\partial\tau}=
-\displaystyle\frac{\hbar^{2}}{2m_{o}}\square\Phi(x,\tau)+V(x)\Phi(x,\tau)-
\displaystyle\frac{m_{o}c^{2}}{2}\Phi(x,\tau). \tag{1.3}
\end{equation}

This is a natural generalization of Schr\"{o}dinger equation in
spacetime since a steady-stay of this equation gives the
Klein-Gordon equation. As a byproduct, a new action $S_{GR}$ with
the general relativity effect is also discussed in (4.6) and as a
special case the formula $E=mc^{2}$ comes out naturally.

\section{The equation of motion with quantum effect}

In this section we shall formulate Einstein's General Relativity and
propose our equation of motion with quantum effect in spacetime.

Following the Geometric Approach (II) to investigate the quantum
effect in General Relativity, we shall employ the geometric
viewpoint to study these two theories. First we give a formulation
of Einstein's General Relativity which consists of two
components :\\[10pt]
\textbf{(GR-I) Geometric Component.}  {\it The space is determined
by Einstein's Field Equation ([We]) :}
\begin{equation}R_{ij}-\displaystyle\frac{R}{2}g_{ij}=-\displaystyle\frac{8\pi
G}{c^{4}}T_{ij}\tag{2.1}\end{equation}
{\it where $R_{ij}$ denotes
the Ricci curvature of the Lorentz metric $g_{ij}$, R the scalar
curvature, $c$ the speed of
light and $T_{ij}$ the energy-momentum tensor.}\\[15pt]

Taking the trace of both sides of the equation (2.1), one obtains
the scalar curvature

\begin{equation}R=\displaystyle\sum\limits_{\footnotesize
i}\displaystyle\frac{8\pi
G}{c^{4}}T_{ij}.\tag{2.2}\end{equation}

This scalar curvature will play an important role in our formulation
of equation of motion from the view point Mach's principle. This
component (GR-I) describes how the geometry of the spacetime forms.
Since Einstein gave his description of the general relativity, a lot
of variants have tried to add some new variables or freedom to
modify Einstein's theory, like Brans-Dicke scalar-tensor theory.
However, it turns out so far that Einstein's theory is the most
elegant and simple one that has passed several
tests.\\[10pt]
\textbf{(GR-II) Physical Component.} {\it The equation of motion for
a photon or massive particle with mass $m_{o}$ moving in the
spacetime is given by the geodesics equation :}
\begin{equation}
\nabla_{\gamma^{'}(\tau)}\gamma^{'}(\tau)=0\tag{2.3}
\end{equation}
{\it where $\nabla$ denotes the covariant derivative associated to
the Lorentz metric tensor $g_{ij}$ for the spacetime and
$\gamma(\tau)$ is the timelike trajectory of the particle under
consideration.}

This Component tells us how a massless or massive particle moves in
spacetime. The equation of motion with quantum effect that we would
like to propose to replace (GR-II) is following
:\\[10pt]
\textbf{(GR-II') Physical Component with Quantum Effect.} {\it For a
massless particle, like photon, the trajectory is given by the
equation :}
\begin{equation}
\nabla_{\gamma^{'}(\tau)}\gamma^{'}(\tau)=\displaystyle\frac{\lambda^{2}c^{2}}{48\pi^{2}}\nabla
 R\tag{2.4}\end{equation}
{\it without external force where $\lambda$ stands for the
wavelength of the massless particle. While for massive particle with
rest mass $m_{o}$, the timelike trajectory is given by the
generalized Newton equation of motion :}
\begin{equation}
F=\nabla_{\gamma^{'}(\tau)}\gamma^{'}(\tau)-\displaystyle\frac{\hbar^{2}}{12m_{o}}\nabla
R\tag{2.5}\end{equation} {\it where $\nabla R$ is the gradient of
the scalar curvature $R$, $\hbar$ the Planck constant and F the
external force.}
 Note that the term
$\frac{\hbar^{2}}{12m_{o}}$$\nabla R$ is the quantum-gravitation
effect in spacetime. It is also worth noticing that vector $\nabla
R(x)$ should be timelike for any point $x$ in spacetime. This can be
seen easily from the fact that if we place a rest particle with mass
$m_{o}$ at $x$, then it should be accelerated in the direction of
$\nabla R(x)$, and hence the vector $\nabla R(x)$ is timelike.
\par Before we explain how this term comes out, we discuss this
term from the viewpoint of Mech's principle. It is well-known that
Mech's principle was considered to be of great importance and value
to Einstein who tried to incorporate it into his general theory of
relativity. However, Einstein's formulation of his general
relativity (GR-I) (GR-II) is only semi-Machian. Since (GR-I)
indicates that the energy-momentum tensor $T_{ij}$ determines the
geometry of spacetime, (GR-I) is distribution can put on the motion
of particles, so it is not Machian. In our formulation (GR-II') we
also include this effect into consideration since the term
$\frac{\hbar^{2}}{12m_{o}}\nabla R$ is related to the
energy-momentum tensor $T_{ij}$ as given in (2.2). Therefore, these
two components (GR-I) and (GR-II) together
give a full Machian theory of General Relativity.\\
\section{ The quantum-gravitation effect : \large$\frac{\hbar^{2}}{12m_{o}}$\normalsize$\nabla R$ and \large$\frac{\lambda^{2}c^{2}}{48\pi^{2}}$\normalsize$\nabla
R$}
\par In this section we shall illustrate theoretically how to
obtain the term $\frac{\hbar^{2}}{12m_{o}}$$\nabla R$ in (2.5) and
why we believe that the generalized Newton's equation of motion
(2.5) in spacetime should be the one we search for when we take
the quantum-gravitation effect into consideration. To do so, we
recall the observations of Dirac and Feynman about the path
integral formulation of Quantum Mechanics in the 3-space
$\mathbb{R}^{3}$.
\par Dirac observed in [Di] the Lagrangian or the action also
played an important role in Quantum Mechanics, especially from the
viewpoint of transformation theory. Feynman pushed further to obtain
his path integral formulation. To be more specific, one considers
the Schr\"{o}dinger equation :
\begin{equation}
i\hbar\displaystyle\frac{\partial\Phi(x,t)}{\partial
t}=-\displaystyle\frac{\hbar^{2}}{2m_{o}}\nabla^{2}\Phi(x,t)+V(x)\Phi(x,t)\tag{3.1}\end{equation}

for a particle with mass $m_{o}$ in $\mathbb{R}^{3}$. Now the
fundamental solution $K(b,a)$ can be written as a path integral :

$$K(b,a) =
\int^b_ae^{\frac{i}{h}S[\gamma(t)]}\mathcal{D}\gamma(t)$$

\normalsize where $S[\gamma(t)]$ denotes the action along the path
$\gamma(t)$ from $a$ to $b$; $a$ denotes the particle at the
position $x_{0}$ and time $t_{0}$, and $b$ at the position $x_{1}$
at time $t_{1}$. In brief, $a=(x_{0},t_{0})$ and $b=(x_{1},t_{1})$.
\begin{itemize}
\item[] The key points that concern us are following :
\begin{itemize}
\item[(1)]The equation of motion can be obtained from the
Euler-Lagrangian equation of the action $S$;

\item[(2)]The action $S$ is inscribed in the short-time amplitude
of the path integral formulation for the fundamental solution of
the Schr\"{o}dinger equation.
\end{itemize}
These together give us a useful duality ([Di] [f]) :
\end{itemize}
\textbf{Classical-Quantum Duality.} {\it The Newton's equation of
motion is called in the short-time amplitude of the path integral
as in the diagram :} {$$F = m_{o}\gamma^{''}(t)
\Longleftrightarrow K(b,a) =
\int^b_ae^{\frac{i}{h}S[\gamma(t)]}\mathcal{D}\gamma(t).$$}
\normalsize {\it where $S[\gamma(t)]$ $=$
{\large$\int^{t_{1}}_{t_{0}}$$\frac{m_{o}}{2}$}$<\gamma^{'}(t),\gamma^{'}(t)>-V(\gamma(t))dt$
and F $=$ $-\nabla V$.}\\
\par Note that it is usually easy to reach the picture of
classical mechanics from quantum mechanics by letting the quantum
parameter $\hbar$ go to zero while it is usually unclear hoe to
formulate the quantum mechanics from classical mechanics. So, this
duality provide such an important guideline.
\par This duality tells us that we can extract the equation of
motion in classical mechanics once we have the path integral formula
for the fundamental solution of the Schr\"{o}dinger equation (3.1).
To get our generalized Newton's equation of motion (2.5) in
spacetime, we shall follow the spirit of Geometric Approach (II) :
first use this duality on the curved spaces (i.e., Riemannian
manifolds) and then employ the Principle of covariance (c.f. [Do])
to obtain the corresponding equation in spacetime. In so doing we
consider a family of time evolution equation :

\begin{equation}
\alpha \displaystyle\frac{\partial\Phi(x,t)}{\partial
t}=\nabla^{2}\Phi(x,t) +W(x)\Phi(x,t) \tag{3.2} \end{equation}
where$\alpha$ is a nonzero complex number with $Re(\alpha)\geq 0.$
When $\alpha$ is a positive real number, this gives a diffusion
(heat) equation and it is related to the theory of Brownian motion.
When $\alpha$ $=$ $-\frac{2im_{o}}{\hbar}$ and $W$ $=$
$-\frac{2m_{o}}{\hbar^{2}}$$V$, this reduces to the Schr\"{o}dinger
equation (3.1).

The natural generalization of this family of time evolution
equations on a Riemannian manifold $M$ is given
by

\begin{equation}
\alpha\displaystyle\frac{\partial\Phi(x,t)}{\partial
t}=\triangle\Phi(x,t) + W(x)\Phi(x,t)\tag{3.3}\end{equation}
where$\triangle$ is the Laplace operator associated to the metric
tensor $g_{ij}$ of $M$.

According to the Classical-Quantum Duality, in order to catch the
Newton's equation of motion on curved spaces $M$ with metric tensor
$g_{ij}$, we need to develop a path integral formulation of the
fundamental solution of the Schr\"{o}dinger equation.
\par Given any nonzero complex number $\alpha$ with $Re(\alpha)\geq
0$, we define the $\alpha$-action, denoted by
$\omega_{\alpha}(\gamma)$, for a curve $\gamma$ : $[t_{0},t_{1}]$
$\rightarrow$ $M$ in the Sobolev space $H^{1,2}([t_{0},t_{1}],M)$,
by

\begin{equation}{\Large\omega_{\alpha}(\gamma)} =
\int^{t_1}_{t_0}-\frac{\alpha}{4}<\gamma^{'}(t),\gamma^{'}(t)> +
\frac{1}{\alpha}W(\gamma(t)) + \frac{1}{6\alpha}R(\gamma(t))dt
\tag{3.4}\end{equation}

where $R(\gamma(t))$ denotes the scalar curvature of $(M,g)$ at the
point $\gamma(t)$. Note that the Sobolev embedding theorem implies
that a curve $\gamma$ in $H^{1,2}([t_{0},t_{1}],M)$ is continuous
since dim$[t_{0},t_{1}]$ $=$ $1$ and $\omega_{\alpha}(\gamma)$ is
well-defined.

When $\alpha$ $=$ $-\frac{2im_{o}}{\hbar}$ and $W$ $=$
$-\frac{2im_{o}}{\hbar^{2}}V$, (3.3) is the Schr\"{o}dinger equation
on curved spaces $(M,(g_{ij}))$ and (3.4) takes the form

\begin{equation}{\Large\omega_{-\frac{2im_{o}}{\hbar}}(\gamma)} =
\frac{i}{h}\int^{t_1}_{t_0}\frac{m_{o}}{2}<\gamma^{'}(t),\gamma^{'}(t)>
- V(\gamma(t)) + \frac{\hbar^{2}}{12m_{o}}R(\gamma(t))dt.
\tag{3.5}\end{equation}
Next we shall discuss some suitable path
subspaces of $H^{1,2}([t_{0},t_{1}],M)$ that will be used to define
the path integrals.\\[10pt]
{\it Definition (Path Spaces $\mathbb{D}_{k}(a,b)$)}. For $k$ $\in$
$\mathbb{N}$, $t_{1}$ $>$ $t_{0}$ $\geqslant$ $0$ and $x$, $y$ $\in$
$M$ we define the path space $\mathbb{D}_{k}(x, t_{0}$, $y$,
$t_{1})$ to be the space of all broken geodesics $\gamma$ :
$[t_{0},t_{1}]$ $\rightarrow$ $M$ from $x$ to $y$ with possible
broken points at $y_{j}$ $:=$
$\gamma(t_{0}+\frac{j(t_{1}-t_{0})}{k})$, $j$ $=$ $1,2,..., k-1$.
Note that $x$ $=$ $y_{0}$ and $y$ $=$ $y_{k}$. Another way to
describe $\gamma$ is that a broken geodesic passing the point
$y_{j}$ at time $t_{0}$ $+$ $\frac{(j-1)(t_{1}-t_{0})}{k}$. The
curve
{\large$\gamma|_{[t_{0}+\frac{(j-1)(t_{1}-t_{0})}{k},~t_{0}+\frac{j(t_{1}-t_{0})}{k}]}$}
is a smooth minimal geodesic from $y_{j-1}$ to $y_{j}$ and thus it
has the speed $||\gamma^{'}(s)||$ $=$
$d(y_{j-1},y_{j})k/(t_{1}-t_{0})$ when $s$ is in
$(t_{0}+\frac{(j-1)(t_{1}-t_{0})}{k},t_{0}+\frac{j(t_{1}-t_{0})}{k})$.
From this description, we know that a broken geodesic $\gamma$ in
$\mathbb{D}_{k}(x, t_{0}$, $y$, $t_{1})$ corresponds canonically to
a ordered set of points $\{y_{j}\}$$^{k-1}_{j=1}$ when the point
$y_{j-1}$ is not in the cut locus of $y_{j}$ (c.f. [BC]). As long as
the integration is concerned, we can always assume this for all
curves $\gamma$ $\in$ $\mathbb{D}_{k}(x, t_{0}$, $y$, $t_{1})$.
Under this assumption, $\mathbb{D}_{k}(x, t_{0}$, $y$, $t_{1})$ has
a natural manifold structure of dimension $n(k-1)$ since it can be
identified with the open submanifold of all points
$(y_{1},y_{2},..., y_{k-1})$ in $M^{k-1}$ such that $y_{j}$ is not a
cut point of $y_{j-1}$ for all $j$ $=$ $1,2,..., k$. Note that in
this way, $\mathbb{D}_{k}(x, t_{0}$, $y$, $t_{1})$ can be viewed as
an open submanifold of $M^{k-1}$ and indeed, it is obtain ed by
taking away a measure zero subset from $M^{k-1}$. We can also set
the starting event point $a$ $=$ $(x,t_{0})$ and the ending event
point $b$ $=$ $(y,t_{1})$ and denote the path spaces
$\mathbb{D}_{k}(x, t_{0}$, $y$, $t_{1})$ briefly by
$\mathbb{D}_{k}$$(a,b)$.

Thus, the path space $\mathbb{D}_{k}(a,b)$ can be endowed with a
canonical measure, denoted by $\mathcal{D}_{k}\gamma$ from the
product manifold $M^{k-1}$ via the correspondence of $\gamma$ and
the ordered set $\{y_{j}\}$$^{k-1}_{j=1}$ viewed as the point
$(y_{1},y_{2},...,y_{k-1})$ in $M^{k-1}$.
\par Let $K_{\alpha}(b,a)$ denote the fundamental solution of the
time evolution equation (3.3) at $a$. In [Wu1,2], we provide a
mathematical theory of the path integral formulation for the
family of time evolution equations (3.3). More specifically, we
prove, under a suitable uniformality condition, the following {
$$
K_{\alpha}(a,b)~=~\lim\limits_{k\rightarrow\infty}\int_{\mathbb{D}_{k}(a,b)}(\frac{k\alpha}{4\pi
t})^{kn/2}e^{\omega_{\alpha_{k}}(\gamma)}\mathcal{D}_{k}\gamma
$$}where $\omega_{\alpha_{k}}(\gamma)$ is given in (3.4) and
$\alpha_{k}$ $=$ $\alpha$ $+$ $k^{-0.001}$ when $Re(\alpha)$ $=$
$0$, $\alpha_{k}$ $=$ $\alpha$ when $Re(\alpha)$ $>$ $0$.

When $\alpha$ is a positive real number, for example 2, this
generalized the Feynman-Kac formula to curved spaces and allows one
to study Brownian motions on those spaces ([CW]). To our current
interest, we take $\alpha$ $=$ $\frac{-2im_{o}}{\hbar}$ and $W$ $=$
$-\frac{2m_{o}}{\hbar^{2}}V$. Thus we obtain the fundamental
solution $K(a,b)$ of the Schr\"{o}dinger equation on curved spaces :
for $s$ $=$ $t_{1}$ $-$ $t_{0}$ {
\[
K(a,b)=\lim\limits_{k\rightarrow\infty}\int_{\mathbb{D}_{
k}(a,b)}(\frac{km_{o}}{2\pi\hbar
si})^{\frac{kn}{2}}{e^{\frac{i}{h}\int^{t_1}_{t_0}\frac{m_{o}}{2}<\gamma^{'}(t),\gamma^{'}(t)>
- V(\gamma(t)) +
\frac{\hbar^{2}}{12m_{o}}R(\gamma(t))dt}}\mathcal{D}_{k}\gamma.
\]}
\par The constant $(\frac{km_{o}}{2\pi\hbar si})^{\frac{kn}{2}}$ is
the normalizing factor. This formula can be put into a physical
notion as

\begin{equation} K(a,b) =
\int^b_ae^{\frac{i}{h}S[\gamma(t)]}\mathcal{D}\gamma(t)\tag{3.6}
\end{equation}
with the action
\begin{equation}
S[\gamma(t)]=\int^{t_1}_{t_0}\displaystyle\frac{m_{o}}{2}<\gamma^{'}(t),\gamma^{'}(t)>
- V(\gamma(t)) + \frac{\hbar^{2}}{12m_{o}}R(\gamma(t))dt.\tag{3.7}
\end{equation}

According to the Classical-Quantum Duality, the Euler-Lagrangian
equation of the action $S[\gamma(t)]$ gives us the corresponding
Newton's equation of motion:

\begin{equation}F=m_{o}\nabla_{\gamma^{'}(\tau)}\gamma^{'}(\tau)-\displaystyle\frac{\hbar^{2}}{12m_{o}}\nabla
R\tag{3.8}\end{equation}
with the external force $F$ $=$ $-\nabla
V$.

So far, the equation (3.8) is a non-relativistic equation of motion
in a curved space, but not in a spacetime. To go from Riemannian
Geometry to General Relativity, we allow the metric tensor $g_{ij}$
to be Lorenz and hence the principle of general covariance and the
principle of minimal gravitational coupling (c.f. [Do]) allow us to
obtain the desired equation of motion (2.5) with quantum-gravitation
effect in spacetime. This completes our search for the equation
(2.5) in (GR-II').

Next we shall use two different ways to derive the equation (2.4) in
(GR-II'). First, we recall two of Einstein's results :
\begin{itemize}
\item[(1)] Einstein's mass-energy formula : $E$ $=$ $m_{o}c^2$,
and \item[(2)] The energy $E$ of massless particles, like photons,
is proportional to its frequency $\nu$ : $E$ $=$ $h\nu$.
\end{itemize}

These two formulas together give us the mass of equivalence, still
denoted by $m_{o}$ for a massless particle :
\begin{equation}m_{o}=\displaystyle\frac{h\nu}{c^{2}}.\tag{3.9}\end{equation}

To obtain the equation (2.4), we rewrite the equation (2.5) by using
(3.9) and get
\begin{equation}\displaystyle\frac{h\nu}{c^{2}}\nabla_{\gamma^{'}(\tau)}\gamma^{'}(\tau)
=\displaystyle\frac{\hbar^{2}c^{2}}{12h\nu}\nabla
R\tag{3.10}\end{equation}
when there is no external force.

Simplify (3.10) to yield
\begin{equation}
\nabla_{\gamma^{'}(\tau)}\gamma^{'}(\tau)=\displaystyle\frac{c^{4}}{48\pi^{2}\nu^{2}}\nabla
R\tag{3.11}\end{equation}
and this is equivalent to the equation
(2.4) in (GR-II') since $\lambda\nu$ $=$ $c$.

Another way to obtain this equation is to use the notion of matter
waves of Prince Louis de Broglie. According to de Broglie, a
particle travelling with a certain momentum $p$ has an associated
matter of wavelength $\lambda$ given by the relation:
\begin{equation}\lambda=\displaystyle\frac{h}{p}.\tag{3.12}\end{equation}
Plugging the formula (3.12) into the equation (2.5), we get
\begin{equation}\displaystyle\frac{h}{\lambda
v}\nabla_{\gamma^{'}(\tau)}\gamma^{'}(\tau)=\displaystyle\frac{\hbar^{2}\lambda
v}{12h}\nabla R\tag{3.13}\end{equation} where $v$ is the speed of
the particle, when there is no external force.

Now for a massless particle, the equation (3.13) gives again the
equation
\begin{equation}\nabla_{\gamma^{'}(\tau)}\gamma^{'}(\tau)=\displaystyle\frac{\lambda^{2}c^{2}}{48\pi^{2}}\nabla
R\tag{3.14}\end{equation}
and this completes our derivation of the
intrinsic force, {\large$\frac{\hbar^{2}}{12m_{o}}$}$\nabla R$ and
{\large$\frac{\lambda^{2}c^{2}}{48\pi^{2}}$}$\nabla R$, due to the
quantum-gravitation effect.
\section{A new relativistic wave equation}

In this section we shall introduce a new relativistic wave equation
in spacetime by taking the notion of physical in spacetime into
consideration. We also following the line of thinking of the
Geometric Approach(II) as described in the introduction. Recall the
Schr\"{o}dinger equation on curved space takes the form:
\begin{equation}
i\hbar\displaystyle\frac{\partial\Phi(x,t)}{\partial
t}=-\displaystyle\frac{\hbar^{2}}{2m_{o}}\triangle\Phi(x,t)+V(x)\Phi(x,t)\tag{4.1}\end{equation}
for a particle with mass $m_{o}$ in a curved space. The equation is
now included the quantum effect and the curved-space effect, but not
the relativity effect. To unify these three effects, we take the
curved spaces to be the spacetime. On one hand, the Riemannian
notion of a 4-dimensional curved space becomes now a 4-dimensions
Lorentz manifold (spacetime) as given by the Einstein's field
equation. Thus a 3-dimensional position $x_{3}$ corresponds to a
4-dimensional spacetime position $x$ $=$ $(ct,x_{3})$. The classical
notion of time is now becoming the notion of proper time in general
relativity. Thus correspondence can be put as
\par~~~~~~~~~~~~Position in a 3-dimensional space : $x_{3}$
\par~~~~~~~~~~~~$\longmapsto$ Position in a 4-dimensional curved space : $x_{4}$
\par~~~~~~~~~~~~$\longmapsto$ Position in a 4-dimensional spacetime : $x$ $=$
$(ct,x)$.\\[5pt]
The correspondence of time for a physical event is
\par~~~~~~~~~~~~Absolute time in a 3-dimensional space
\par~~~~~~~~~~~~$\longmapsto$ Absolute time in a 4-dimensional
curved space
\par~~~~~~~~~~~~$\longmapsto$ Proper time in a 4-dimensional spacetime\\[5pt]
Under these correspondences, we can transfer the Schr\"{o}dinger
equation on curved space into a spacetime :
\begin{equation}i\hbar\displaystyle\frac{\partial\Phi(x,\tau)}{\partial\tau}=-\displaystyle\frac{\hbar^{2}}{2m_{o}}\square\Phi(x,\tau)+V(x)\Phi(x,\tau)+\omega\Phi(x,\tau)
\tag{4.2}\end{equation}
 where $\square$ denotes the d'Alembertian
operator associated to the Lorentz metric tensor $g_{ij}$ and
$\omega$ is a universal constant related to the relativity effect
that need to be determined. The d'Alembertian operator in the
Minkowski space is given by
\par~~~~~~~~~~~~~~~~~~~~$\square$ =
$\displaystyle\frac{1}{c^{2}}$$\displaystyle\frac{\partial^{2}}{\partial
t^{2}}$ $-$ $\displaystyle\sum\limits_{\footnotesize
i=1}^{\footnotesize 3}$$\displaystyle\frac{\partial^{2}}{\partial
x^{2}_{i}}$.\\[1pt]
\par To find the correct value of relativistic constant $\omega$,
we consider the special case that the Hamiltonian is conservative
and thus this reduces to the equation
\begin{equation}
-\displaystyle\frac{\hbar^{2}}{2m_{o}}\square$$\Phi(x,\tau)+V(x)\Phi(x)+\omega\Phi(x)=0.\tag{4.3}\end{equation}
When the potential $V$ vanishes, this equation should be equivalent
to the classical Klein-Gordon equation:
\begin{equation}\square$$\Phi(x)+(\displaystyle\frac{m_{o}c}{\hbar})^2\Phi(x)=0.\tag{4.4}\end{equation}
Thus the relativistic constant is $\omega$ $=$
$-\displaystyle\frac{m_{o}c^{2}}{2}$. Pluging this constant into
the equation (4.2) we obtain\\
\textbf {A relativistic wave equation.}
\begin{equation}i\hbar\displaystyle\frac{\partial\Phi(x,\tau)}{\partial\tau}=-\displaystyle\frac{\hbar^{2}}{2m_{o}}
\square\Phi(x,\tau)+V(x)\Phi(x,\tau)-\displaystyle\frac{m_{o}c}{2}\Phi(x,\tau).\tag{4.5}\end{equation}

In view of this equation and the generalized Newton's equation of
motion (2.5), the correct action $S_{GR}(\gamma)$ for a path
$\gamma$ from [0,1] into the spacetime should take the form :
\begin{equation}
{S_{GR}(\gamma)=\int^{t_1}_{t_0}\frac{m_{o}}{2}<\gamma^{'}(t),\gamma^{'}(t)>
+ \frac{\hbar^{2}}{12m_{o}}R(\gamma(s)) +
\frac{m_{o}c^{2}}{2}ds}.\tag{4.6}
\end{equation}
In particular, in the Minkowski space a rest particle with mass
$m_{o}$ will have energy :
\par~~~~~~~~~~~~$\displaystyle\frac{m_{o}}{2}$$<\gamma^{'}(s),\gamma^{'}(s)>$
$+$ $\displaystyle\frac{\hbar^{2}}{12m_{o}}R(\gamma(s))$ $+$
$\displaystyle\frac{m_{o}c^{2}}{2}$\\
\par~~~~~~~~~~~~~~~~~~~~~~~~~~$=$
$\displaystyle\frac{m_{o}}{2}(c^{2}-0)$ $+$ $0$ $+$
$\displaystyle\frac{m_{o}c^{2}}{2}$\\
\par~~~~~~~~~~~~~~~~~~~~~~~~~~$=$ $m_{o}c^{2}$.\\[5pt]
This gives the well known formula : $E$ $=$ $m_{o}c^{2}$. In
general, we have $E=m_0c^2 +
\frac{\hbar^2}{12m_0}R$ in a spacetime.\\

\section*{Acknowledgment} The author would like to think C.-R. Lee for
discussions concerning this work, especially Dirac's viewpoint of
Lagrangian played in Quantum Mechanics, and M.-H. Chi for her
interest in this work.

\clearpage
\begin{center}
{\sc References}
\end{center}
\begin{itemize}
\item[{\bf\footnotesize [[BC]]}]\footnotesize R. Bishop and R.L.
Crittendon, {\it Geometry of manifolds}, Academic Press, New York,
1974.

\item[{\bf\footnotesize [[BD]]}]\footnotesize J. D. Bjorken and S.
D. Drell, {\it Relativistic Quantum Mechanics}, McGraw-Hill Book
Company.

\item[{\bf\footnotesize [[Cg]]}]\footnotesize K. S. Cheng, {\it
Quantization of a general Dynamical system by Feynman's path
integration formulation}, Jour. Math. Phys. {\bf 13} (1972),
1723-1726.

\item[{\bf\footnotesize [[Cl]]}]\footnotesize I. Chavel, {\it
Eigenvalues in Riemannian geometry}, Academic Press, New York,
1984.

\item[{\bf\footnotesize [[CW]]}]\footnotesize M.-H. Chi and J.-Y.
Wu, {\it Brownian motion on curved spaces}, in preparation.

\item[{\bf\footnotesize [[DE]]}]\footnotesize C. DeWitt-Morette
and K.D. Elworthy, {\it A stepping stone to stochastic analysis.
IN: New stochastic methods in physics.}, Physics Reports {\bf
77(3)} (1981), 121-382.

\item[{\bf\footnotesize [[De]]}]\footnotesize B.S. DeWitt, Rev.
Mod. Phys. {\bf 29} (1957), 337.

\item[{\bf\footnotesize [[Di]]}]\footnotesize P.A.M. Dirac, {\it
The lagrangian in Quantum Mechanics}, Physikalische Zeitschrift
der Sowjetunion, Band 3, Heft 1 (1933), 64-70.

\item[{\bf\footnotesize [[Do]]}]\footnotesize R. D'Inverno, {\it
Introducing Einstein's Relativity}, Clarendon Press, Oxford, 1992.

\item[{\bf\footnotesize [[ET]]}]\footnotesize K. D. Elworthy and
A. Truman, {\it Classical mechanics, the diffusion (heat) equation
and the Schr\"{o}dinger equation}, J. Math : Phys. {\bf 22-10}
(1981), 2144-2166.

\item[{\bf\footnotesize [[F]]}]\footnotesize R.P. Feynman, {\it
Space-time approach to non-relativistic equation mechanics}, Rev.
Med. Pyhs. {\bf 20} (1948), 267.

\item[{\bf\footnotesize [[FH]]}]\footnotesize R.P. Feynman and
A.R. Hibbs, {\it Quantum Mechanics and Path integrals},
McGraw-Hill Publishing Company, New York, 1965.

\item[{\bf\footnotesize [[G]]}]\footnotesize Y. Gliklikh, {\it
Global Analysis in Mathematical Physics : Geometric and Stochastic
Methods}, Springer, 1997.

\item[{\bf\footnotesize [[K1]]}]\footnotesize H. Kleinert, {\it
Path integrals in Quantum Mechanics Statistics and Polymer
Physics, 2nd edition}, World Scientific, Singapore, 1995.

\item[{\bf\footnotesize [[K2]]}]\footnotesize H. Kleinert, Mod.
Phys. Lett. A {\bf 4} (1989), 2329.

\item[{\bf\footnotesize [[N]]}]\footnotesize {\it Feynman
integrals and the Schr\"{o}dinger equation}, J. Math. Phys {\bf 5}
(1964), 332-343.

\item[{\bf\footnotesize [[R]]}]\footnotesize G. Roepstorff, {\it
Path integral approach to Quantum Processes}, Springer-Verlag,
1994.

\item[{\bf\footnotesize [[Sa]]}]\footnotesize Sabbata (ed.), {\it
Quantum Mechanics in Curved Spacetime}, Plenum Press, 1990.

\item[{\bf\footnotesize [[Sw]]}]\footnotesize M. Swanson, {\it
Path Integrals and Quantum Processes}, Academic Press1992.

\item[{\bf\footnotesize [[U]]}]\footnotesize A. Unterberger, {\it
Quantization, Symmetries and Relativity}, Contemporary
Mathematics, vol 214 : Perspectives on Quantization (1996),
169-195.

\item[{\bf\footnotesize [[We]]}]\footnotesize S. Weinberg, {\it
Gravitation and Cosmology : Principles and applications of the
general theory of relativity}, John Wiley $\&$ Sons, Inc., 1972.

\item[{\bf\footnotesize [[Wu1]]}]\footnotesize J.-Y. Wu, {\it A
mathematical background for path integrals on curved spaces},
submitted, 1-14.

\item[{\bf\footnotesize [[Wu2]]}]\footnotesize J.-Y. Wu, {\it
Curvature, bounded cohomology and path integrals}, to appear in
the Proceeding of ICCM'98, Beijing.

\item[{\bf\footnotesize [[Wu3]]}]\footnotesize J.-Y. Wu, {\it
Gravitational Rainbow}, NCCU Math technical report No. JYWU
1999-4.

\item[{\bf\footnotesize [[Wu4]]}]J.-Y. Wu, {\it The Evaporation of
Black Holes}, NCCU Math technical report No. JYWU 1999-5.

\end{itemize}

\end{document}